\documentclass[10pt,twocolumn,letterpaper]{article}

\usepackage{cvpr}
\usepackage{times}
\usepackage{epsfig}
\usepackage{graphicx}
\usepackage{amsmath}
\usepackage{amssymb}
\usepackage{upgreek}
\usepackage{caption} 
\usepackage[breaklinks=true,bookmarks=false]{hyperref}

\cvprfinalcopy % *** Uncomment this line for the final submission

\def\cvprPaperID{26} % *** Enter the CVPR Paper ID here

\begin{document}

%%%%%%%%% TITLE
\title{A Hybrid Approach Between Adversarial Generative Networks and Actor-Critic Policy Gradient  for Low Rate High-Resolution Image Compression}

\author{Nicol\'o Savioli\\
 \href{}{Imperial College London, UK}\\
{\tt\small nsavioli@ic.ac.uk}}

\maketitle
%\thispagestyle{empty}

%%%%%%%%% ABSTRACT
\begin{abstract}
Image compression is an essential approach for decreasing the size in bytes of the image without deteriorating the quality of it. Typically, classic algorithms are used but recently deep-learning has been successfully applied.
In this work, is presented a deep super-resolution work-flow for image compression that maps low-resolution JPEG image to the high-resolution. The pipeline consists of two components: first, an encoder-decoder neural network learns how to transform the downsampling JPEG  images to high resolution. Second, a combination between Generative Adversarial Networks (GANs) and reinforcement learning  Actor-Critic (A3C) loss pushes the encoder-decoder to indirectly maximize High Peak Signal-to-Noise Ratio (PSNR). Although PSNR is a fully differentiable metric, this work opens the doors to new solutions for maximizing non-differential metrics through an end-to-end approach between encoder-decoder networks and reinforcement learning policy gradient methods.
\end{abstract}

%%%%%%%%% BODY TEXT
\section{Introduction}

Image compression with deep learning systems is an active area of research that recently has  becomes very compelling respect to the modern natural images codecs as JPEG2000, \cite{nla.cat-vn72353}, BPG \cite{BPG} WebP currently developed by $Google^{\tiny{\textregistered}}$ \cite{WebP}.
The new deep learning methods are based on an auto-encoder architecture where the features maps, generate from a Convolutional Neural Networks (CNN) encoder, are passed through a quantizer to create a binary representation of them, and subsequently given in input to a CNN decoder for the final reconstruction.    
In this view, several encoders and decoders models have been suggested as a ResNet \cite{DBLP:journals/corr/HeZRS15} style network with the parametric rectified linear units (PReLU) \cite{Liu_2018_CVPR_Workshops}, generative approach build on GANs  \cite{Agustsson_2018_CVPR_Workshops} or with a innovative hybrid networks made with Gated Recurrent Units (GRUs) and ResNet \cite{Toderici_2017_CVPR}.
In contrast, this paper proposes a super-resolution approach, build on a modifying version of SRGAN \cite{DBLP:conf/cvpr/LedigTHCCAATTWS17}, where downsampling JPEG images are converted at High Resolution (HR) images.
Hence, in order to improve the final PSNR results, a Reinforcement Learning (RL) approach is used to indirectly maximize the PSNR function with an A3C policy \cite{DBLP:journals/corr/MnihBMGLHSK16} end-to-end joined with SRGAN.
The main contributions of this works are: (i) Propose a compression pipeline based on JPEG image downsampling combined with a super-resolution deep network. (ii) Suggest a new way for maximizing not differentiable metrics through RL. However, even if the PSNR metric is a fully differentiable function, the proposed method could be used in future applications for non-euclidean distance such as in the Dynamic Time Warping (DTW) algorithms \cite{Keogh:2000:SUD:347090.347153}.

\section{Methods}

In this section is given a more formal description of the suggested system which includes: the network architecture and the losses used for training.

\subsection{Network Architecture}

\begin{figure*}[ht]
\centering
 \includegraphics[width=19.5cm,height=5.5cm]{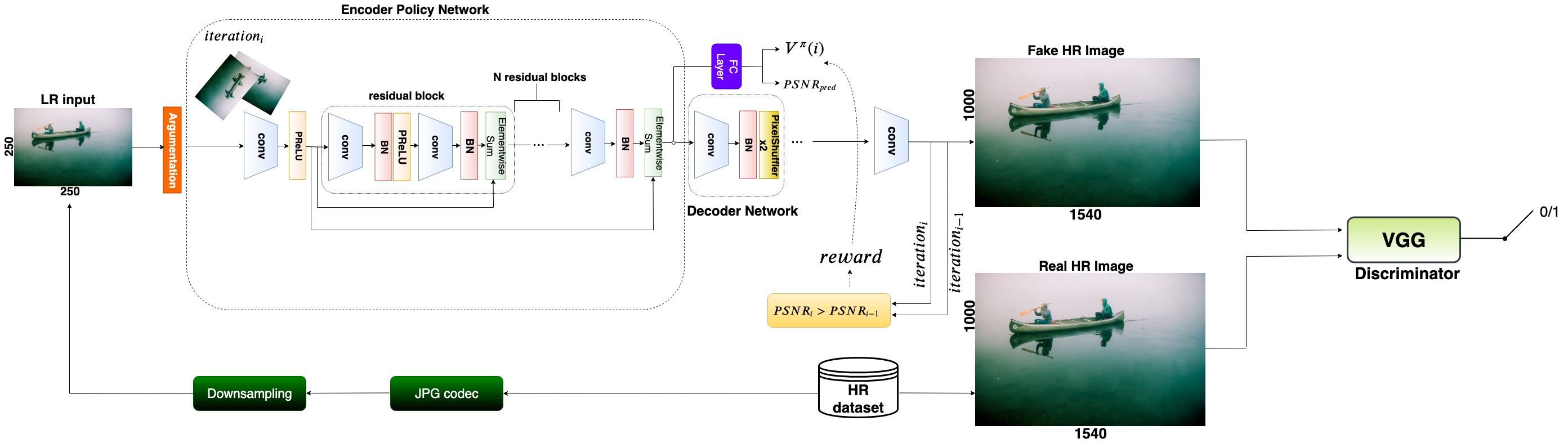}
   \caption{The figure shows the proposed RL-SRGAN model composed by encoder, decoder and discriminator networks.
The objective of this model is to map the compressed JPEG Low Resolution (LR) Image to the HR. 
The encoder can be seen as: (i) an RL policy network able to increase its HR prediction through the indirect maximization of the PSNR at each $i$ training iterations. (ii) A GANs, where the discriminator (i.e VGG network) push the decoder to produce images similar to the original HR ground truth.}
  \label{fig:model}
\end{figure*}

The architecture consists of three main blocks: encoder, decoder and a discriminator (Figure \ref{fig:model}).
The  $I^{LR}$  is  the low-resolution (LR) input image (i.e compressed with a JPEG encoder) of size $rW \times rH \times C$ ( i.e with C color channels and W, H the image width and height); where a bicubic downsampling operation with factor $r$ is applied. While the output is an HR image defined as $I^{HR}$.

\subsubsection{Encoder}

The encoder is basically a ResNet  \cite{DBLP:journals/corr/HeZRS15}, where the first convolution block has a kernel size of  $9\times 9$ and $64$ Feature Maps (FM) with a $ParametricReLU$ activation function. 
Then, five Residual Blocks (RB) are stacked together.  
Each of those RB consists of two convolution layers with kernel size $3\times3$ and $64$ FM followed by Batch-Normalisation (BN) and $ParametricReLU$.  
After that, a final convolution block of $3\times3$ and $64$ FM are repeated.
However, the encoder is also joint with a fully connected layer and, at each $i$ training iterations, produces an action prediction of the actual PSNR; together with a value function $V^{\pi}(I^{HR}(i))$ (i.e explained in \ref{par:rloss} section).

\subsubsection{Decoder}

The decoder is fundamentally another deep network that allows increasing the resolution of the output encoder with eight subpixel layers \cite{DBLP:journals/corr/AitkenLTCWS17}.

\subsubsection{Discriminator}

The encoder, joint with the decoder, define a generator $H(\cdot)_{\theta}$, where $\theta = [w_{L}; b_{L}]$ are the weight and biases parameters for each L-layers for the specific network.
A third network $D(\cdot)_{\theta}$, called discriminator, is also optimized concurrently with $H(\cdot)_{\theta}$ for solving the following adversarial min-max problem: 

\begin{equation}
  \begin{aligned}
      l_{GAN}^{HR} = min_{\theta} \thinspace \thinspace max_{\theta} E_{I^{HR} \sim {p_{train}(I^{HR})}}[log(D(I^{HR}))]  + \\ 
      +  E_{I^{LR} \sim {p_{H}}(I^{LR})}[log(1-D(H(I^{LR})))]
 \label{eq:one}
  \end{aligned}
\end{equation}

The idea behind  $l_{GAN}^{HR}$  loss is to train a generative model $H(\cdot)_{\theta}$ to fool $D(\cdot)_{\theta}$. 
Indeed, the discriminator is trained to distinguish super-resolution images $I^{HR}$, generated by $H(\cdot)_{\theta}$, from those of the training dataset.
In this way, the discriminator is increasingly struggled to distinguish the  $I^{HR}$ images (generated by $H(\cdot)_{\theta}$) from the real ones and consequently driving the generator to produce results closer to the HR training images. 
Then, in the proposed model, the discriminator $D(\cdot)_{\theta}$ is parameterized through a VGG network with LeakyReLU activation $(\alpha = 0.2)$ without max-pooling.

\subsection{Loss function}

The accurate definition of the loss function is crucial for the performance of the $H(\cdot)_{\theta}$ generator. 
Here, the paragraph is logically divided into three losses: the SRGAN loss, the RL loss, and the proposed loss.

\subsubsection{SRGAN loss}

The SRGAN loss is determined as a combination of three other separate losses: MSE loss, VGG loss, and GANs loss.
Where the MSE loss is defined as:

\begin{equation} 
l_{MSE}^{HR}= \frac{1}{WH} \sum_{x=1}^{W}  \sum_{y=1}^{H} (I_{x,y}^{HR} - H_{\theta}(I^{LR})_{x,y})^{2}
\end{equation}

It represents the most utilized loss in super-resolution methods but remarkably sensitive to high-frequency peak with smooth textures \cite{DBLP:conf/cvpr/LedigTHCCAATTWS17} .
For this reason, is used a VGG loss \cite{DBLP:conf/cvpr/LedigTHCCAATTWS17}  based on the ReLU activation function of a 19 layer VGG (defined here as $\Omega(\cdot)$) network:

\begin{equation} 
l_{VGG}^{HR}= \frac{1}{WH} \sum_{x=1}^{W}  \sum_{y=1}^{H} (\Omega(I^{HR})_{x,y} - \Omega(H_{\theta}(I^{LR})_{x,y})^{2}
\end{equation}

Where $W$ and $H$ are the dimension of $I^{HR}$ image in the MSE loss. 
Whilst, for the VGG loss, they are the $\Omega(\cdot)$ output FM dimensions.
While the GANs loss is previously defined in the equation \ref{eq:one}. 
Finally, the total SRGAN loss is determined as: 

\begin{equation} 
l_{SRGAN}^{HR}=  l_{MSE}^{HR} +1e-3 \times l_{GAN}^{HR} + 6e-3 \times  l_{VGG}^{HR}
\end{equation}

\subsubsection{RL loss}   \label{par:rloss}

The aim of RL loss is to indirectly maximize the PSNR through an actor-critic approach  \cite{DBLP:journals/corr/MnihBMGLHSK16}. 
Given $Q^{\pi}(I^{LR},PSNR_{pred})$ a map between the low resolution input $I^{LR}$ and the current PSNR value prediction $PSNR_{pred}$ (see fig. \ref{fig:model}).
Thence, at each $i$ training iterations, is calculated the reward value as a threshold between the previous $PSNR$ at iteration $i-1$ and that one to iteration $i$ as follows:

\begin{equation} 
    r(i)= 
\begin{cases}
    1 ,& \text{if} \thinspace \thinspace \thinspace PSNR_{i} > PSNR_{i-1}\\
    0,              & \text{otherwise}
\end{cases}
 \label{eq:reward}
\end{equation}

where the $PSNR(\cdot)$ function is defined as: 

\begin{equation} 
  \begin{aligned}
PSNR = 20 \cdot \log_{10} (MAX_{I})- \\ 10 \cdot \log_{10}(\frac{1}{mn} \sum_{l=0}^{m-1} \sum_{j=0}^{n-1} [I^{HR}(l,j)-I^{HR}_{gt}(l,j)]^{2})
  \end{aligned}
 \label{eq:psnr}
\end{equation}

The $MAX_{I}$ is the maximum pixel value of the HR image, $I^{HR}$ is the output encoder HR image, while $I^{HR}_{gt}$ is the corresponding HR ground truth for each pixel $(l,j)$ at $m \times n$ HR size.
The reward (eq. \ref{eq:reward}), actually depends on the $PSNR_{pred}$ action taken by the  policy for two main reasons: (i) during the training process the $PSNR_{pred}$ becomes an optimal estimator of the decoder output $I^{HR}$ (used in \ref{eq:psnr}). (ii) The latent space between the encoder and the fully connected layer is the same and share equal policy information.
Thus, all the rewards are accumulated every $k$ training steps through the following return function:

\begin{equation} 
R(i) = \sum_{k=0}^{\infty} \gamma^{k} r(i+k)
\end{equation}

where $\gamma \in (0,1]$ is a discount factor.
Therefore, is possible to define the $Q^{\pi}(\cdot)$ function as an expectation of $R(k)$ given the input $I^{LR}$ and $PSNR_{pred}$.

\begin{equation} 
Q^{\pi}(I^{LR},PSNR_{pred})= E[R(i)|I^{LR}(i)=I^{LR},PSNR_{pred}]
\end{equation}

To notice, the encoder, together with the fully connected layer, become the policy network  $\pi(PSNR_{pred}|I^{LR}(i);\theta_{H})$.
This policy network is parametrized by the standard $REINFORCE$ method on the $\theta$ encoder parameters with the following gradient direction:

\begin{equation} 
\nabla_{\theta} \log \pi(PSNR_{pred}|I^{LR}(i);\theta) \cdot R(i)
\end{equation}

It can be consider an unbiased estimation of $\nabla_{\theta}  \cdot E[R(i)]$.
Especially, to reduce the variance of this evaluation (and keeping it unbiased) is desirable to subtract, from the return function, a baseline $V(I^{LR}(i))$ called value function. 
The total policy agent gradient is given by:

\begin{equation} 
l_{\pi}^{HR}  =  \log \pi(PSNR_{pred}|I^{LR}(i),\theta) \cdot (R(i)-V(I^{LR}(i)))
 \label{eq:six}
\end{equation}

\begin{figure}[ht]
\centering
\includegraphics[width=\linewidth]{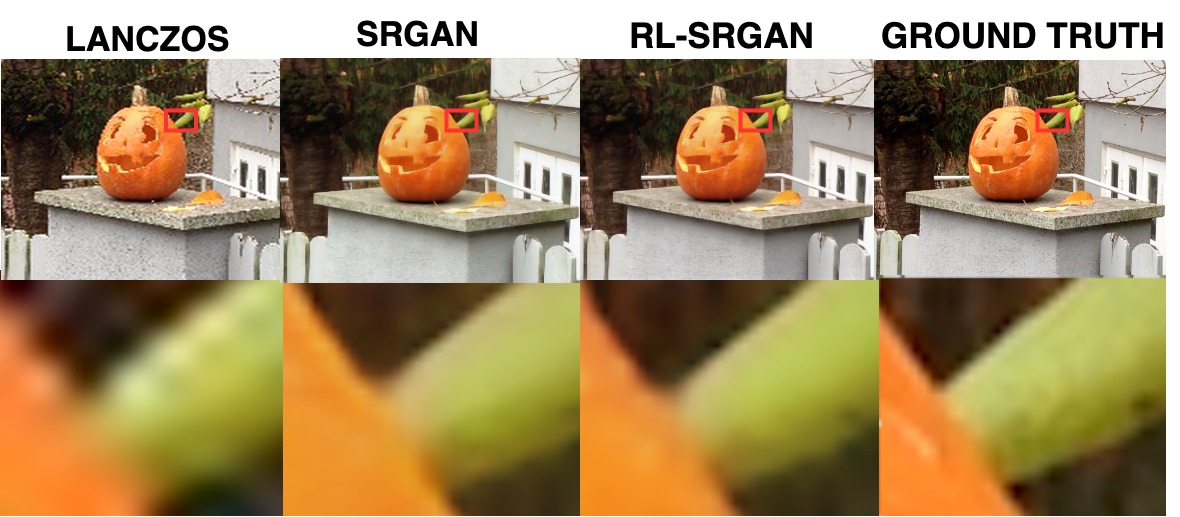}
   \caption{The above figure shows the results for RL-SRGAN compared to SRGAN, LANCZOS, and the original ground truth.
As we can see, LANCZOS simply destroys the edges producing artifacts on the global image. Whilst, SRGAN forms noticeable chromatic aberration (i.e transition from orange to yellow color) near the edges. Even though, the RL-SRGAN holds the color uniform with net outline details nearby to the edges; analogous to the original ground truth.}
  \label{fig:result}
\end{figure}

The term $R(i)-V(I^{LR}(i))$ can be considered a estimation of the advantage to predict $PSNR_{pred}$ for a given $I^{LR}(i)$ input. 
Consequently, a learnable evaluation of the value function is used:  $V(I^{LR}(i)) \approx V^{\pi}(I^{LR}(i))$. 
This approach, is further called generative actor-critic \cite{DBLP:journals/corr/MnihBMGLHSK16} becouse the $PSNR_{pred}$ prediction is the actor while the baseline $V^{\pi}(I^{LR}(i))$ is its critic.
The RL loss is then calculated as:

\begin{equation} 
l_{RL}^{HR} =  5e-3*\sum (R(i)-V^{\pi}(I^{LR}(i)))^2 - l_{\pi}^{HR}  
\end{equation}
\subsubsection{Proposed loss}

The Proposed Loss (PL) combines both SRGAN loss and RL loss.
After every $k$ step (i.e due to the rewards accumulation process at each $i$ training iterations), the $l_{RL}^{HR}$ is added on $l_{SRGAN}^{HR}$.

\begin{equation} 
    l_{PL}^{HR}= 
\begin{cases}
     l_{SRGAN}^{HR} + l_{RL}^{HR} ,& \text{if} \thinspace \thinspace \thinspace  k = i \\
     l_{SRGAN}^{HR}  , & \text{otherwise}
\end{cases}
\end{equation}

\subsection{Experiments and Results}

In this section is evaluate the method suggested.
The dataset used is the CLIC compression dataset \cite{clic} correspondingly divided
in the train, valid and test sets. The train has $1634$ HR images, valid $102$  and test $330$.
The evaluation metrics used are the PSNR and MS-SSIM \cite{Wang03multi-scalestructural}
for both valid and test. 
An ADAM optimizer is used with a learning rate of $1e-3$ within $22876$ model iterations until convergence.
The Reinforcement Learning SRGAN (RL-SRGAN) is compared with the SRGAN model work \cite{DBLP:conf/cvpr/LedigTHCCAATTWS17} and the Lanczos resampling (i.e a smooth interpolation through a convolution between the $I^{LR}$ image and a stretched $sinc(\cdot)$ function).
Finally, the table \ref{table:one} highlights that the PSNR difference between LANCZOS upsampling and RL-SRGAN is 0.9, while of 0.19 with SRGAN; whereas the MS-SSIM remains constant between RL-SRGAN and SRGAN for the validation set. This also shows a better accuracy for the RL-SRGAN model.  
While, for the tests,  RL-SRGAN  achieve $20.06$ of PSNR and $0.7503$ of MS-SSIM. 
Furthermore, the compression rate for the validation set images is 3.812.623 bytes respect 362.236.068 bytes of original HR dataset. While for the test set images is 5.228.411 bytes in contrast with the 5.882.850.012 bytes of the original one. That makes the method a good trade-off between compression capacity and acceptable PSNR.

\begin{table}[]
\begin{tabular}{|l|l|l|}
\hline
\textit{\textbf{Methods}} & \textit{\textbf{PSNR}} & \textit{\textbf{MS-SSIM}} \\ \hline
RL-SRGAN                  & \textbf{22.34}         & \textbf{0.783}            \\ \hline
SRGAN                     & 22.15                  & 0.780                     \\ \hline
LANCZOS                 & 21.44                  & 0.760                     \\ \hline
\end{tabular}
  \caption{The table shows the PSNR and MS-SSIM results obtained in the validation set for the proposed rl-srgan method with srgan and lanczos upsampling.}
 \label{table:one}
\end{table}

\subsection{Discussion}

A modified version of SRGAN is suggested where an A3C method is joined with GANs. 
Sadly, the proposed method has strong limitations due to the drastic downsampling of the input JPEG image.
This downsampling causes loss of information, difficult to recover from the super-resolution network, which leads to lower results in PSNR and MS-SSIM on the test set (i.e $20.06$ and $0.7503$ respectively). Despite, the results (table  \ref{table:one}) emphasize slight improvement performances for RL-SRGAN related within SRGAN and a baseline LANCZOS upsampling filter. 
However, the proposed method compresses all test files in a parsimonious way respect to the challenge methods.
Indeed, the total dimension of the compression test set is of 5236870 bytes respect to 15748677 bytes of CLIC 2019 winner. 
Finally, a new method for maximizing non-differentiable functions is here suggested through deep reinforcement learning technique. 

{\small
\bibliographystyle{unsrt}
\bibliography{egbib}

\begin{thebibliography}{10}

\bibitem{nla.cat-vn72353}
David~S. Taubman and Michael~W. Marcellin.
\newblock {\em JPEG2000 : image compression fundamentals, standards, and
  practice / David S. Taubman, Michael W. Marcellin}.
\newblock Kluwer Academic Publishers Boston, 2002.

\bibitem{BPG}
Fabrice bellard. bpg image format.
\newblock \url{https://bellard.org/bpg}.

\bibitem{WebP}
Webp image format.
\newblock \url{https://developers.google.com/speed/webp}.

\bibitem{DBLP:journals/corr/HeZRS15}
Kaiming He, Xiangyu Zhang, Shaoqing Ren, and Jian Sun.
\newblock Deep residual learning for image recognition.
\newblock {\em CoRR}, abs/1512.03385, 2015.

\bibitem{Liu_2018_CVPR_Workshops}
Haojie Liu, Tong Chen, Qiu Shen, Tao Yue, and Zhan Ma.
\newblock Deep image compression via end-to-end learning.
\newblock In {\em The IEEE Conference on Computer Vision and Pattern
  Recognition (CVPR) Workshops}, June 2018.

\bibitem{Agustsson_2018_CVPR_Workshops}
Eirikur Agustsson, Michael Tschannen, Fabian Mentzer, Radu Timofte, and Luc
  Van~Gool.
\newblock Extreme learned image compression with gans.
\newblock In {\em The IEEE Conference on Computer Vision and Pattern
  Recognition (CVPR) Workshops}, June 2018.

\bibitem{Toderici_2017_CVPR}
George Toderici, Damien Vincent, Nick Johnston, Sung Jin~Hwang, David Minnen,
  Joel Shor, and Michele Covell.
\newblock Full resolution image compression with recurrent neural networks.
\newblock In {\em The IEEE Conference on Computer Vision and Pattern
  Recognition (CVPR)}, July 2017.

\bibitem{DBLP:conf/cvpr/LedigTHCCAATTWS17}
Christian Ledig, Lucas Theis, Ferenc Huszar, Jose Caballero, Andrew Cunningham,
  Alejandro Acosta, Andrew~P. Aitken, Alykhan Tejani, Johannes Totz, Zehan
  Wang, and Wenzhe Shi.
\newblock Photo-realistic single image super-resolution using a generative
  adversarial network.
\newblock In {\em 2017 {IEEE} Conference on Computer Vision and Pattern
  Recognition, {CVPR} 2017, Honolulu, HI, USA, July 21-26, 2017}, pages
  105--114, 2017.

\bibitem{DBLP:journals/corr/MnihBMGLHSK16}
Volodymyr Mnih, Adri{\`{a}}~Puigdom{\`{e}}nech Badia, Mehdi Mirza, Alex Graves,
  Timothy~P. Lillicrap, Tim Harley, David Silver, and Koray Kavukcuoglu.
\newblock Asynchronous methods for deep reinforcement learning.
\newblock {\em CoRR}, abs/1602.01783, 2016.

\bibitem{Keogh:2000:SUD:347090.347153}
Eamonn~J. Keogh and Michael~J. Pazzani.
\newblock Scaling up dynamic time warping for datamining applications.
\newblock In {\em Proceedings of the Sixth ACM SIGKDD International Conference
  on Knowledge Discovery and Data Mining}, pages 285--289. ACM, 2000.

\bibitem{DBLP:journals/corr/AitkenLTCWS17}
Andrew~P. Aitken, Christian Ledig, Lucas Theis, Jose Caballero, Zehan Wang, and
  Wenzhe Shi.
\newblock Checkerboard artifact free sub-pixel convolution: {A} note on
  sub-pixel convolution, resize convolution and convolution resize.
\newblock {\em CoRR}, abs/1707.02937, 2017.

\bibitem{clic}
Workshop and challenge on learned image compression (clic).
\newblock \url{http://www.compression.cc/}.

\bibitem{Wang03multi-scalestructural}
Zhou Wang, Eero~P. Simoncelli, and Alan~C. Bovik.
\newblock Multi-scale structural similarity for image quality assessment.
\newblock pages 1398--1402, 2003.

\end{thebibliography}
}

\end{document}